------------------------------------------------------------------------
\documentstyle[preprint,aps]{revtex}

\begin{document}
\title{Quantum scattering in one dimension}
\author{Vania E Barlette$^1$, Marcelo M Leite$^2$ and Sadhan K
Adhikari$^3$}
\address{$^1$Centro Universit\'ario Franciscano, Rua dos Andradas
1614, 97010-032 Santa
Maria,
RS, Brazil\\
$^2$Departamento de F\'isica, Instituto Tecnol\'ogico de Aeronautica,
Centro T\'ecnico Aeroespacial, 12228-900 S\~ao Jos\'e dos Campos, SP,
Brazil\\
 $^3$Instituto de F\'{\i}sica Te\'orica, 
Universidade Estadual Paulista\\
01.405-900 S\~ao Paulo, SP, Brazil\\
}
\date{\today}
\maketitle

\begin{abstract} 

A self-contained discussion of nonrelativistic quantum scattering is
presented in the case of central potentials in one space dimension, which 
will facilitate the  understanding of the  more complex scattering theory 
in two and three dimensions. The present discussion illustrates in a
simple way the concept of  partial-wave decomposition, phase shift,  
optical
theorem and  effective-range expansion.

\end{abstract}



\newpage

\section{Introduction}

There has been great interest in studying tunneling phenomena in a finite
superlattice \cite{1}, which essentially deals with quantum scattering in
one dimension. It has also been emphasized \cite{2} that a study of quantum
scattering in one dimension is important for the study of Levinson's theorem
and virial coefficient. Moreover, one-dimensional quantum scattering
continues
as an active line of research \cite{2a}. Because of these recent interests
we present a comprehensive description of quantum scattering in one
dimension in close analogy with the two- \cite{3} and three-dimensional
cases \cite{5}. Apart from these interests in research, the study of
one-dimensional scattering is also interesting from a pedagogical point of
view. In a one dimensional treatment one does not need special mathematical
functions, such as the Bessel's functions, while still retaining
sufficient
complexity to illustrate many of the physical processes which occur
in two and three dimensions. Hence the present discussion of one
dimensional scattering is expected to  assist the understanding of
the more complex two and three dimensional scattering problems.

A
self-contained discussion of two-dimensional scattering has appeared in the
literature \cite{3}. It is realized that the scattering amplitude which
arises from the usual asymptotic form of the scattering wave function in 
three dimensions is not
the most satisfactory one in two dimensions and a modified scattering
amplitude has been proposed in this case. However, this modification in the
case of two dimensions can not be extended to one-dimensional scattering in
a straightforward way, basically because in one dimension there are only two
discrete scattering directions: forward and backward along a line. This
requires special techniques in one dimension in distinction with two or
three dimension where an infinity of scattering directions are permitted
characterized by continuous angular variable(s).

It is because of the above subtlety of the one-dimensional scattering
problem we
present a complete discussion of this case. Here we present a new relation
between the asymptotic wave function and the scattering amplitude with
desirable analytic properties and consequently, present a discussion of
one-dimensional scattering which should be considered complimentary to the
discussion of two-dimensional scattering of Ref. \cite{3}.

In Sec. 2 we present a wave-function description of scattering with a
convenient partial wave analysis. Section 3 illustrates the the
formulation of effective-range expansion. A brief summary is presented in
Sec. 4. 

\section{Wave-function description}

In one dimension the scattering wave function $\psi _{k}^{(+)}(x)$ at
position $x$ satisfies the Schr\"{o}dinger equation 
\begin{equation}
\left[ -\frac{\hbar ^{2}}{2m}\frac{d^{2}}{dx^{2}}+V(x)\right] \psi
_{k}^{(+)}(x)=E\psi _{k}^{(+)}(x)  \label{1}
\end{equation}
where $V(x)$ is a centrally-symmetric finite-range potential satisfying $
V(x)=0, x_0\equiv |x|>R$ and $V(x)=V(-x)$, where  $m$ is the reduced mass,
$E\equiv
\hbar ^{2}k^{2}/(2m)$ the energy and $k$ the wave number.

The asymptotic form of the wave function is taken as 
\begin{equation}
\lim_{x_0\rightarrow \infty }\psi _{k}^{(+)}(x)\rightarrow \exp
(ikx)+\frac{i}{
k}f_{k}^{(+)}(\epsilon )\exp (ikx_0)  \label{4}
\end{equation}
where  $\epsilon =x/x_0,$ $f_{k}^{(+)}(\epsilon )$ is the
scattering
amplitude, $
\exp (ikx)$ the incident plane wave and $\exp (ikx_0)$ the scattered
outgoing
wave. There are two discrete directions in one dimension characterized by
signs of $x$ and given by $\epsilon =\pm 1$ in contrast to infinite
possibility of 
scattering angles in two and three dimensions. The forward (backward)
direction is given by the positive (negative) sign. The discrete
differential cross sections in these two directions are given by 
\begin{equation}
\sigma _{\epsilon }=\frac{1}{k^{2}}|f_{k}^{(+)}(\epsilon )|^{2}  \label{5}
\end{equation}
and the total cross section by 
\begin{equation}
\sigma _{\text{tot}}=\sum_{\epsilon }\sigma _{\epsilon }=\frac{1}{k^{2}}
\left[|f_{k}^{(+)}(+)|^{2}+|f_{k}^{(+)}(-)|^{2}\right].  \label{6}
\end{equation}
The discrete sum in  (\ref{6}) is to be compared with integrals over
continuous angle variables in two and three dimensions. The two differential
cross sections (\ref{5}) are  the usual reflection and transmission
probabilities. An extra factor of $i$ is introduced in the outgoing wave
part of  (\ref{4}). This will have the advantage of leading to an optical
theorem in close analogy with the three-dimensional case \cite{5} as we
shall see in the following.

In order to define the partial-wave projection of the wave function and
partial-wave phase shifts, we consider the following parametrization of the
asymptotic wave function 
\begin{equation}
\lim_{x_0\to \infty}\psi _{k}^{(+)}(x)=\sum_{L=0}^{1}\epsilon
^{L}A_{L}\cos
\left[ kx_0+\frac{L\pi }{2}
+\delta _{L}(k^2)\right]  \label{7}
\end{equation}
where $\delta _{L}$ is the scattering phase shift for the $L$th wave and $
A_{L}$ is an unknown coefficient. In contrast to the infinite
number of partial waves
in two and three dimensions, two partial waves are sufficient in this case.
Also, one has the following partial-wave projection for the incident wave 
\begin{equation}
\exp (ikx)=\text{cos}(kx_0)+i\epsilon \sin (kx_0).  \label{8}
\end{equation}
Consistency between  (\ref{4}) and (\ref{7}) yields 
$
A_{L}=(-i)^{L}\exp (i\delta _{L})  \label{9}
$
and 
\begin{equation}
f_{k}^{(+)}(\epsilon )=k\sum_{L=0}^{1}\epsilon ^{L}\exp (i\delta _{L})\sin
(\delta _{L}).  \label{10}
\end{equation}

Now one naturally defines the partial-wave amplitudes 
\begin{equation}
f_{L}=k\exp (i\delta _{L})\sin (\delta _{L})  \label{11}
\end{equation}
in terms of phase shift $\delta _{L}$ so that the total cross section of 
(\ref{6}) becomes 
\begin{equation}
\sigma _{\text{tot}}=2\sum_{L=0}^{1}\sin ^{2}(\delta _{L})  \label{12}
\end{equation}
and satisfies the following optical theorem 
\begin{equation}
\sigma _{\text{tot}}=\frac{2}{k}\Im f_{k}^{(+)}(+),  \label{13}
\end{equation}
in close analogy with the three-dimensional case, where $\Im $ is the
imaginary part. Had we not introduced $i$ in the asymptotic form (\ref{4}),
this optical theorem would involve the real part of the forward scattering
amplitude in place of the imaginary part. This factor of $i$ has no
consequence on the observables.

In partial waves $L=0,1$, the Schr\"odinger equation (\ref{1}) becomes 
\begin{equation}  \label{14}
\left[ -\frac{\hbar^2}{2m}\frac{d^2}{dx_0^2}
+V(x_0)\right]\psi_{k,L}^{(+)}(x_0) =
E\psi_{k,L}^{(+)}(x_0)
\end{equation}
where $x_0\equiv |x|$ by definition is positive. The two solutions $-$
symmetric and antisymmetric, corresponding to $L=0$ and 1, respectively $-$
are constructed from  (\ref{14}) using the asymptotic condition (\ref{7}).

\section{Effective-range expansion}

Next we illustrate the formulation of the last section by developing an
effective-range expansion in both partial waves $L=0$ and 1 in close
analogy with three dimensions. There are some specific difficulties in the
two dimensional case which we shall not consider here \cite{6}. In both
partial waves the radial Schr\"{o}dinger equation can be written as
\begin{equation} \left[ \frac{d^{2}}{dx_0^{2}}+k^{2}-U(x_0)\right] \psi
_{k,L}^{(+)}(x_0)=0, \label{40} \end{equation} where
$U(x_0)=2mV(x_0)/\hbar^2$. The $L=0$ wave involves the symmetric solution
and the $L=1$ wave the antisymmetric solution. The solution of (\ref{40})
behaves like $\psi (x_0)=\chi (kx_0)$ near $x_0=0$ and $\psi (x_0)=\chi
[kx_0+\delta _{L}(k^2)]$ in the asymptotic region, $x_0 \rightarrow \infty
$. The function $\chi $ is a sine function for $L=1$ and a cosine one for
$L=0$. This compact notation will simplify the treatment of the
effective-range expansion. The treatment of the $L=1$ solution in one
dimension is essentially identical with that of the S-wave potential
scattering in three dimensions. The similarity between the present
one-dimensional treatment for $L=1$ and the S-wave potential scattering in
three dimensions suggests that the usual form of \ Levinson's theorem in
three dimensions will be valid in this case, e.g., $\delta_1
(0)-\delta_1(\infty)
=n\pi ,$
where $n$ \ is the number of $L=1$ bound states in one dimension \cite{2}.

We consider two solutions $u_{1}$ and $u_{2}$ at two energies $k_{1}^{2}$
and $k_{2}^{2}$ of the Schr\"{o}dinger equation (\ref{40}). These solutions
are normalized such that asymptotically 
\begin{equation}
\lim_{x_0\rightarrow \infty }u_{i}(x_0)\rightarrow \frac{\chi
[k_{i}x_0+\delta
_{L}(k_i^2)]}{\chi [\delta _{L}(k_i^2)]}  \label{41}
\end{equation}
where $i=1$ and $2$ refer to the two energies $k_{1}^{2}$ and $k_{2}^{2}$,
respectively. From the two equations (\ref{40}) satisfied by the functions $
u_{1}$ and $u_2$ we readily obtain as in the three-dimensional case 
\begin{eqnarray}  \label{42}
\left[ u_{2}(x_0)u_{1}^{\prime }(x_0)-u_{1}(x_0)u_{2}^{\prime
}(x_0)\right]
_{0}^{R}=(k_{2}^{2}-k_{1}^{2})\int_{0}^{R}u_{1}(x_0)u_{2}(x_0)dx_0
\end{eqnarray}
where $R$ is an arbitrary radial distance and the prime on the function
$u$
denotes derivative with respect to $x_0$.

Next we consider two free-particle solutions 
\begin{equation}
v_{i}(x_0)=\frac{\chi [k_{i}x_0+\delta _{L}(k_i^2)]}{\chi [\delta
_{L}(k_i^2)]} \label{43}
\end{equation}
of  (\ref{40}) obtained by putting $U(x_0)=0$. Then we have 
\begin{eqnarray}  \label{44}
\left[ v_{2}(x_0)v_{1}^{\prime }(x_0)-v_{1}(x_0)v_{2}^{\prime
}(x_0)\right]
_{0}^{R}=(k_{2}^{2}-k_{1}^{2})\int_{0}^{R}v_{1}(x_0)v_{2}(x_0)dx_0.
\end{eqnarray}

With these general equations first we specialize to the case of $L=1$.
Subtracting  (\ref{44}) from (\ref{42}), using  (\ref{41}) and (\ref
{43}) and the conditions $u_{1}(0)=u_{2}(0)=0$ and letting $R$ go to $\infty 
$, one gets 
\begin{equation}
k_{2}\cot \delta _{1}(k_2^2)=k_{1}\cot \delta
_{1}(k_1^2)+(k_{2}^{2}-k_{1}^{2})\int_{0}^{\infty
}[v_{1}(x_0)v_{2}(x_0)-u_{1}(x_0)u_{2}(x_0)]dx_0.  \label{47}
\end{equation}
It is convenient to define the $L=1$ scattering length $a_{1}$ in analogy
with three dimensions by 
\begin{equation}
-\frac{1}{a_{1}}=\lim_{k\rightarrow 0}k\cot \delta _{1}(k^2).  \label{48}
\end{equation}
Now letting $k_{1}=0$ and denoting $k_{2}=k$ in  (\ref{47}) we have 
\begin{equation}
k\cot \delta _{1}(k^2)=-\frac{1}{a_{1}}+k^{2}\int_{0}^{\infty
}[v_{0}(x_0)v_{k}(x_0)-u_{0}(x_0)u_{k}(x_0)]dx_0  \label{49}
\end{equation}
where the suffix 0 or $k$ on the wave function refers to the energy. Next
making a power-series expansion of the integral in (\ref{49}), we get the
one-dimensional effective-range expansion 
\begin{equation}
k\cot \delta _{1}(k^2)=-\frac{1}{a_{1}}+\frac{1}{2}r_{1}k^{2}+O(k^{4})+...
\label{50}
\end{equation}
with 
\begin{equation}
r_{1}=2\int_{0}^{\infty }[v_{0}^{2}(x_0)-u_{0}^{2}(x_0)]dx_0  \label{51}
\end{equation}
where $u_{0}$ and $v_{0}$ are the zero-energy solutions and $r_{1}$ the
effective range for $L=1$.

The meaning of scattering length and effective range becomes more explicit
if we consider the symmetric one-dimensional square well defined by
$U(x_0)=0$
for $x_0>R$ and $U(x_0)=-\beta _{0}^{2}$ for $x_0<R$. The solution of the
Schr\"{o}dinger equation in this (antisymmetric) case is given by 
\begin{eqnarray}
u(x_0) &=&B_{1}\sin (\beta x_0),\quad x_0<R \\
&=&A_{1}\sin [kx_0+\delta _{1}(k^2)],\quad x_0>R
\end{eqnarray}
where $\beta ^{2}=k^{2}+\beta _{0}^{2}$.
By matching the log derivative at $x_0=R$ one gets $k\cot [kR+\delta
_{1}(k^2)]=\beta \cot (\beta R)$.
This
condition can be rewritten as 
\begin{eqnarray}  \label{52}
k\cot \delta _{1}(k^2)=\frac{k^{2}\tan (\beta R)\tan (kR)+k\beta }{k\tan
(\beta
R)-\beta \tan (kR)}.
\end{eqnarray}
A straightforward low-energy expansion of (\ref{52}) and comparison with 
(\ref{50}) yield 
\begin{eqnarray}
a_{1}=R-\frac{1}{\beta _{0}}\tan (\beta _{0}R)
\end{eqnarray}
\begin{eqnarray}  \label{53}
r_{1}=2R-2\frac{R^{2}}{a_{1}}+\frac{2R^{3}}{3a_{1}^{2}}+\left( 1-\frac{R}{
a_{1}}\right) ^{2}\left( \frac{1}{\beta _{0}\tan (\beta _{0}R)}-\frac{R}{
\sin ^{2}(\beta _{0}R)}\right).
\end{eqnarray}

In this case the first bound state appears for $\beta_0R>\pi/2$ and 
a new bound state appears as $\beta_0R $ crosses $(2n+1)\pi/2$
where $n$ is a positive integer including 0. So the system will have $n$
bound states for $(2n-1)\pi/2 < \beta_0R <(2n+1)\pi/2$. The scattering
length $a_1$ tends to infinity as $\beta_0R \to (2n+1)\pi/2$, which denotes
the appearance of a new bound state. Also, as a new bound state appears, $
a_1 \to \infty$ and $r_1$ of  (\ref{53}) tends to $R$ the range of
square well. This is why the name effective range is given to $r_1$.

Next we consider the case of $L=0$. Although the $L=1$ case discussed
above is quite similar to the three dimensional case for $L=0$, the $L=0$
case described below has no three-dimensional analogue. Nevertheless we
describe this case for the sake of completeness.  
 Subtracting  (\ref{44}) from (\ref{42}
), using  (\ref{41}) and (\ref{43}) and the conditions $u_{1}^{\prime
}(0)=u_{2}^{\prime }(0)=0$ and letting $R$ go to $\infty $, one gets 
\begin{equation}
k_{2}\tan \delta _{0}(k^2_2)=k_{1}\tan \delta
_{0}(k_1^2)-(k_{2}^{2}-k_{1}^{2})\int_{0}^{\infty
}[v_{1}(x_0)v_{2}(x_0)-u_{1}(x_0)u_{2}(x_0)]dx_0.  \label{56}
\end{equation}
It is convenient to define the $L=0$ scattering length $a_{0}$ by 
\begin{equation}
\frac{1}{a_{0}}=\lim_{k\rightarrow 0}k\tan \delta _{0}(k^2).
\end{equation}
Now letting $k_{1}=0$ and denoting $k_{2}=k$ in  (\ref{56}) we have 
\begin{equation}
k\tan \delta _{0}(k^2)=\frac{1}{a_{0}}-k^{2}\int_{0}^{\infty
}[v_{0}(x_0)v_{k}(x_0)-u_{0}(x_0)u_{k}(x_0)]dx_0  \label{58}
\end{equation}
where the suffix 0 or $k$ on the wave function refers to the energy. A
power-series expansion of  (\ref{58}) in energy leads to 
\begin{equation}\label{xx}
k\tan \delta _{0}(k^2)=\frac{1}{a_{0}}+\frac{1}{2}r_{0}k^{2}+O(k^{4})+...
\end{equation}
where the effective range is given by 
\begin{equation}
r_{0}=-2\int_{0}^{\infty }[v_{0}^{2}(x_0)-u_{0}^{2}(x_0)]dx_0.
\end{equation}

The solution of the Schr\"{o}dinger equation in this case for the
one-dimensional square well defined by $U(x_0)=0$ for $x_0>R$ and
$U(x_0)=-\beta
_{0}^{2}$ for $x_0<R$ is given by 
\begin{eqnarray}
u(x_0) &=&B_{0}\cos (\beta x_0),\quad x_0<R \\
&=&A_{0}\cos [kx_0+\delta _{0}(k^2)],\quad x_0>R
\end{eqnarray}
By matching the log derivative at $x_0=R$ one gets $k\tan [kR+\delta
_{0}(k^2)]=\beta \tan (\beta R)$ . This condition can be rewritten as 
\begin{equation}\label{kk}
k\tan \delta _{0}(k^2)=\frac{k\beta \tan (\beta R)-k^{2}\tan (kR)}{k+\beta
\tan
(\beta R)\tan (kR)}.
\end{equation}
A straightforward low-energy expansion of (\ref{kk}) and comparison with 
(\ref{xx}) yield 
\begin{equation}
a_{0}=R+\frac{1}{\beta _{0}\tan (\beta _{0}R)}
\end{equation}
\begin{equation}\label{57}
r_{0}=2R-2\frac{R^{2}}{a_{0}}+\frac{2R^{3}}{3a_{0}^{2}}-\left( 1-\frac{R}{
a_{0}}\right) ^{2}\left( \frac{\tan (\beta _{0}R)}{\beta _{0}}+\frac{R}{\cos
^{2}(\beta _{0}R)}\right) 
\end{equation}

In this case there is at least one bound state for all values of
$\beta_0R$ and  a new bound state appears as $\beta _{0}R$ crosses $n\pi $
where $n$ is a positive integer including 0. So the system will have $n$
bound states for $(n-1)\pi <\beta _{0}R<n\pi $. The scattering length $a_{0}$
tends to infinity as $\beta _{0}R\rightarrow n\pi $ denoting the appearance
of a new bound state. Also, as a new bound state appears, $a_{0}\rightarrow
\infty $, and $r_{0}$ of  (\ref{57}) tends to $R$ the range of square
well. This is why the name effective range is given to $r_{0} $ in this case.

\section{Summary}

In this paper we have generalized the standard treatments of two- and
three-dimensional scattering to the case of one-dimensional scattering. We
have introduced the concept of scattering amplitude, partial wave, phase
shift, optical theorem  and effective-range expansion in close analogy
with three-dimensional scattering. In this case there are two partial
waves: $L=0 $ and 1. 
The
quantum mechanical one-dimensional scattering is mathematically a much
simpler problem
compared to that in two and three dimensions. 
The
present discussion of one-dimensional scattering deviod of the usual
mathematical complexity encountered in two and three dimensions 
will help one to  understand easily the different physical concepts  
of general scattering theory.

The work is supported in part by the Conselho Nacional de Desenvolvimento -
Cient\'{i}fico e Tecnol\'{o}gico, Funda\c{c}\~{a}o de Amparo \`{a} Pesquisa
do Estado de S\~{a}o Paulo, and Finan\-ciadora de Estu\-dos e Projetos of
Brazil.

\end{document}